\documentclass{article}

\usepackage{graphicx}
\usepackage{subfigure}
\usepackage{stfloats}
\usepackage{geometry}
\usepackage{multicol}
\usepackage{hyperref}
\usepackage{caption}
\usepackage{authblk}
\usepackage{hyperref}

\hypersetup{
    colorlinks=true,
    linkcolor=blue,
    urlcolor=blue
}
\title{\Large\bfseries Application of Deep Learning Methods Combined with Physical Background in Wide Field of View Imaging Atmospheric Cherenkov Telescopes}
\author[1]{\normalsize Ao-Yan Cheng}
\author[1]{Hao Cai\thanks{Corresponding author: \href{mailto:hcai@whu.edu.cn}{hcai@whu.edu.cn}}}
\author[2]{Shi Chen}
\author[3]{Tian-Lu Chen}
\author[1]{Xiang Dong}
\author[3]{You-Liang Feng}
\author[3]{Qi Gao}
\author[4]{Quan-Bu Gou}
\author[4,5]{Yi-Qing Guo}
\author[4,5]{Hong-Bo Hu}
\author[6]{Ming-Ming Kang}
\author[3]{Hai-Jin Li}
\author[4]{Chen Liu}
\author[3]{Mao-Yuan Liu}
\author[4,5]{Wei Liu}
\author[4]{Fang-Sheng Min}
\author[1]{Chu-Cheng Pan}
\author[4]{Bing-Qiang Qiao}
\author[7]{Xiang-Li Qian}
\author[7]{Hui-Ying Sun}
\author[1]{Yu-Chang Sun}
\author[1]{Ao-Bo Wang}
\author[7]{Xu Wang}
\author[8]{Zhen Wang}
\author[9]{Guang-Guang Xin}
\author[4,10]{Yu-Hua Yao}
\author[11]{Qiang Yuan}
\author[11]{Yi Zhang}

\affil[1]{\small School of Physics and Technology, Wuhan University, Wuhan 430072, People's Republic of China}
\affil[2]{School of Physics and Astronomy, Yunnan University, Yunnan 650091, China}
\affil[3]{The Key Laboratory of Cosmic Rays (Tibet University), Ministry of Education, Lhasa 850000, Tibet, People’s Republic of China}
\affil[4]{Key Laboratory of Particle Astrophysics, Institute of High Energy Physics, Chinese Academy of Sciences, Beijing 100049, People's Republic of China}
\affil[5]{University of Chinese Academy of Sciences, 19 A Yuquan Road, Shijingshan District, Beijing, 100049, China}
\affil[6]{College of Physics, Sichuan University, Chengdu, 610064, China}
\affil[7]{School of Intelligent Engineering, Shandong Management University, Jinan, 250357, China}
\affil[8]{Tsung-Dao Lee Institute, Shanghai Jiao Tong University, 200240 Shanghai, China}
\affil[9]{Suzhou Aerospace Information Research Institute, Suzhou 215123, China}
\affil[10]{College of Physics, Chongqing University, No.55 Daxuecheng South Road, High-tech District, Chongqing, 401331, China}
\affil[11]{Key Laboratory of Dark Matter and Space Astronomy, Purple Mountain Observatory, Chinese Academy of Sciences, Nanjing 210008, China}
\date{}

\begin{document}
\maketitle
\begin{abstract}
The HADAR experiment, which will be constructed in Tibet, China, combines the wide-angle advantages of traditional EAS array detectors with the high sensitivity advantages of focused Cherenkov detectors. Its physics objective is to observe transient sources such as gamma-ray bursts and counterparts of gravitational waves. The aim of this study is to utilize the latest AI technology to enhance the sensitivity of the HADAR experiment. We have built training datasets and models with distinctive creativity by incorporating relevant physical theories for various applications. They are able to determine the kind, energy, and direction of incident particles after careful design. We have obtained a background identification accuracy of 98.6\%, a relative energy reconstruction error of 10.0\%, and an angular resolution of 0.22-degrees in a test dataset at 10 TeV. These findings demonstrate the enormous potential for enhancing the precision and dependability of detector data analysis in astrophysical research. Thanks to deep learning techniques, the HADAR experiment's observational sensitivity to the Crab Nebula has surpassed that of MAGIC and H.E.S.S. at energies below 0.5 TeV and remains competitive with conventional narrow-field Cherenkov telescopes at higher energies. Additionally, our experiment offers a fresh approach to dealing with strongly connected scattered data.

\textbf{Keywords: } VHE Gamma-Ray Astronomy, HADAR, Deep Learning, Convolutional Neural Networks
\end{abstract}	

\begin{multicols}{2}
\section{Introduction}
The investigation of ultra-high-energy gamma-ray astronomy \cite{RN75} is pivotal for understanding a range of extreme astrophysical phenomena. These ultra-high-energy gamma rays predominantly originate from highly active galactic nuclei, cataclysmic supernova explosions, neutron stars, and black holes. Importantly, these observations form an empirical cornerstone for addressing some of the most compelling scientific questions, including the detection of dark matter and the identification of cosmic ray sources.

Space observatories like the Fermi Gamma-ray Space Telescope \cite{RN74} need long observation times to gather statistically relevant data because high-energy gamma-ray flux decays quickly and has a power-law distribution. Consequently, larger-area ground detectors are required for these high-energy gamma rays.

The Earth's atmosphere becomes into a substantial interaction medium for high-energy photons in the energy range above GeV \cite{RN88}, effectively preventing their transmission. In this regime, high-energy $\gamma$-rays interact with Earth's upper atmospheric layers \cite{RN95}, leading to the generation of electron-positron $e^- e^+$ pairs, which in turn instigate electromagnetic cascades. Within these cascades, the relativistic electrons and positrons produce concentrated beams of Cherenkov radiation. These beams serve as the principal observational targets for our specialized ground-based detection systems, enabling nuanced studies of the high-energy universe.

Gamma-ray astronomy has become one of the frontier disciplines for studying the universe's highest energy astrophysical phenomena over the last few decades. The Cherenkov Telescope Array (CTA) is an international initiative that seeks to better understand high-energy gamma rays in the universe by identifying atmospheric Cherenkov radiation \cite{RN89}. The 20 GeV to 300 TeV energy range covered by CTA is intended to close the observational gap currently present in this energy range. A group of five Cherenkov telescopes called the High Energy Stereoscopic System (HESS) is now operational in Namibia and is used to observe cosmic TeV energy rays. It offers crucial information for comprehending gamma-ray sources \cite{RN90} like pulsars and supernova remnants \cite{RN91}. The Major Atmospheric Gamma Imaging Cherenkov (MAGIC) telescopes, two telescopes situated on La Palma in the Canary Islands, are another notable project. Through the detection of Cherenkov radiation from cosmic gamma-ray sources \cite{RN85}, they investigate the physical characteristics and origins of cosmic rays \cite{RN93}. These telescopes with a small field of view are primarily employed to make accurate observations of known high-energy astrophysical objects, providing important knowledge and observational data for our comprehensive understanding of these extreme cosmological occurrences.

The measurable signals from many extreme celestial occurrences, however, only last for brief periods because of their intrinsically transitory character. There is an urgent demand for wide-field detectors that can instantly capture high-energy particle signals across a wide spatial range since large detector arrays require time to recalibrate their directed reception. Wide-field Cherenkov imaging telescopes are especially useful in this situation.

The HADAR experiment exemplifies this technology \cite{RN86}, comprising an array of imaging atmospheric Cherenkov refractive telescopes anchored in atmospheric Cherenkov principles. Positioned at an impressive altitude of 4300 meters in Yangbajing, Tibet (N 30.0848, E 90.5522), the experiment incorporates four water lenses, each spanning a diameter of 5 meters, strategically placed at the vertices of a square with side lengths of 100 meters. These lenses, ensconced within a hemispherical glass shell of 5 meters in diameter, are housed atop an 8-meter diameter steel structure tank. This tank, rising to a height of 7 meters, is brimming with high-purity water, while its base is adorned with cameras outfitted with photomultiplier tubes. Upon the interception of Cherenkov radiation, the quartet of detectors meticulously logs the charge deposition on the photomultiplier tubes within a predefined temporal frame. All collected data are subsequently archived in an external storage apparatus.

To effectively detect very-high-energy (VHE) $\gamma$-ray sources, analytical methodologies must execute several critical tasks:
\begin{enumerate}
\item Background Suppression: The identification of distinct shape features within detector images is pivotal. This allows for the isolation of target $\gamma$-rays from the overwhelmingly abundant background of cosmic rays, which are predominantly composed of protons.
\item Energy Reconstruction: The accurate estimation of the original energy of incident particles is achieved by correlating variables such as deposited charge and the relative spatial coordinates within the detectors.
\item Direction Reconstruction: Leveraging the stereoscopic images acquired from the detectors, it is crucial to reconstruct the axis direction of the resultant particle shower. This, in turn, facilitates the estimation of the original direction from which the incident particles emanated.
\end{enumerate}

Deep learning techniques are increasingly becoming integral to the data analysis frameworks employed in Cherenkov telescope experiments. A variety of computational algorithms, particularly those based on convolutional neural networks (CNNs), have shown unparalleled success in addressing the multifaceted analytical challenges endemic to this scientific domain. Prominent initiatives such as H.E.S.S.\cite{RN73} and CTA\cite{RN81}\cite{RN82} have adeptly harnessed the robust capabilities of CNNs to scrutinize their observational datasets, yielding significant advancements. These technological strides corroborate the findings delineated in the present study, reinforcing the compelling case for the widespread utility and robustness of deep learning methodologies within this specialized field of research.

The organization of this paper's structure is as follows: The use of deep learning is briefly introduced in Section 2 along with our data production process. A more thorough overview will be provided in Sections 3 to 5, and the pertinent data will be used as the basis for the training of background suppression, energy reconstruction, and incident direction models. The modeling calculations of the flux and observational sensitivity of the Crab Nebula at various energies are covered in Section 6. Finally, Section 7 will offer a thorough summary of the material presented.

\section{Deep-Learning Approaches for HADAR Data}
Deep Learning is an important branch of Artificial Intelligence (AI) and has emerged as a growing field in recent years. With the significant increase in computational power, especially with GPU chips, real-time computation of large parallel data (e.g., high-dimensional matrices) has become possible. This has led to the convergence of deep learning towards various data analysis industries, simplifying tasks that were previously difficult for humans to perform. As a type of artificial intelligence, deep learning methods simulate the multi-layer neural network structure of the human brain to solve problems. Essentially similar to human learning, deep learning methods continuously update network parameters through a step-by-step understanding of data. This process ultimately allows the network to extract as many useful features as possible from the inputs and obtain model predictions through neural network computations.

In our work, we have adopted Convolutional Neural Networks (CNNs). As an excellent model for image recognition, CNNs can spontaneously extract local relationships at different positions within the same dimension and across different dimensions based on the input signals, ultimately yielding optimal results. Fundamentally, a CNN is composed of multiple layers: specifically, it consists of a series of convolutional-pooling layers connected to fully connected layers. Due to this layered structure, errors from one layer get propagated to the next, potentially leading to an exponential growth of errors. To mitigate this, we use a specialized CNN, known as a Residual Convolutional Neural Network. The advantage of this network is that each layer receives not only the processed signals from the previous layer but also the original signals from the previous layer. This significantly enhances the model's ability to fit the data.

By leveraging the power of CNNs, specifically Residual CNNs, we aim to tackle the three main challenges in detecting VHE gamma-ray sources: background suppression, energy reconstruction, and direction reconstruction. Our experimental results, which we will detail in subsequent sections, demonstrate the effectiveness of deep learning methods in these high-stakes, complex analytical tasks.

Throughout the research process, all datasets used for training the neural networks were generated using the widely-adopted CORSIKA Monte Carlo program \cite{RN63} \cite{RN64}. In the simulations, the set altitude was 4300 $m$, corresponding to an atmospheric depth of 606 g/cm$^2$. The geomagnetic coordinates were set to the location of Yangbajing in Tibet. The simulated primary cosmic ray particles included both gamma rays (serving as signal) and protons (serving as background), with energy ranges spanning from 20 GeV to 10 TeV. The incident zenith angle was set to 20$^\circ$, and the azimuthal angle ranged from 0 to 360$^\circ$. All events were uniformly scattered within a circle centered on the HADAR array, with a radius of 400 $m$. After the shower simulation, the response of the HADAR detector was simulated using appropriate software packages, and corresponding telescope images were generated.

To implement the model incorporating the associated algorithms, we utilized the deep learning framework based on PyTorch. Training and testing were carried out on a machine equipped with 2 NVIDIA GeForce GTX 3090 GPUs and a high-performance computing cluster furnished with 4 NVIDIA V100 Tensor Core GPUs, achieving nearly identical results on both platforms.

For diverse research tasks, we have curated specialized datasets, adopted a range of neural network architectures, and employed various loss functions. Detailed descriptions of these components will be thoroughly discussed in the subsequent three sections, where we will also articulate the underlying rationale for these choices. Our codebase, named 'GPLearn,' is publicly available on GitHub. This package features modular functionalities, enabling users to tailor module parameters to specific needs. With GPLearn, readers have the capability not only to reproduce the results presented in this paper using our supplied datasets but also to adapt the code for actual detector data analysis. We anticipate that these contributions will find broad applicability in the field of astrophysics.

To successfully execute GPLearn, ensure that you have both Python and the compatible version of PyTorch installed on your system. To install GPLearn, simply clone the repository by executing the following command in your terminal:
\begin{center}
git clone \href{https://github.com/caihao/gpLearn}{https://github.com/caihao/gpLearn.git}
\end{center}
This will install the most up-to-date package in your current directory. For detailed usage guidelines and instructions, kindly consult the appropriate tutorials.

\section{Background Suppression}
The flux of cosmic rays typically exceeds that of high-energy gamma rays by several orders of magnitude, posing a significant challenge for detection. Consequently, the Cherenkov radiation generated by these high-energy cosmic rays acts as a predominant source of background noise. Given that approximately 90\% of cosmic rays are proton constituents, this study primarily focuses on evaluating the influence of protons on photon detection. Particularly problematic are proton signals with energies threefold greater than those of photon signals, as these are especially difficult to distinguish. To tackle this challenge, we have conducted simulations involving particles with a range of energies, aiming to effectively segregate the target photon signals from the interfering proton background.

\begin{figure}[htbp]
\centering
\subfigure{
	\label{Fig.7.1}
	\includegraphics[width=0.45\textwidth]{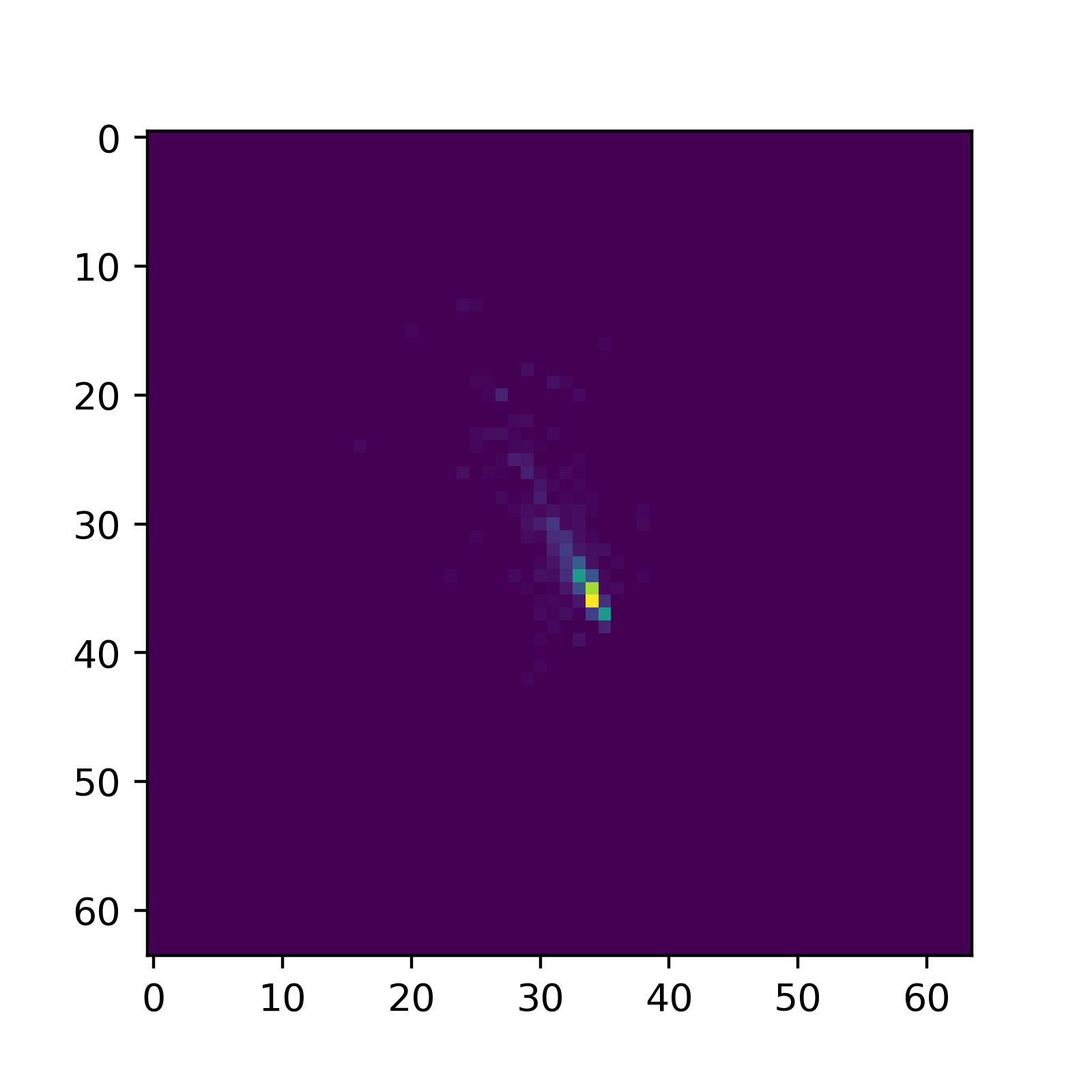}
}
\subfigure{
	\label{Fig.7.2}
	\includegraphics[width=0.45\textwidth]{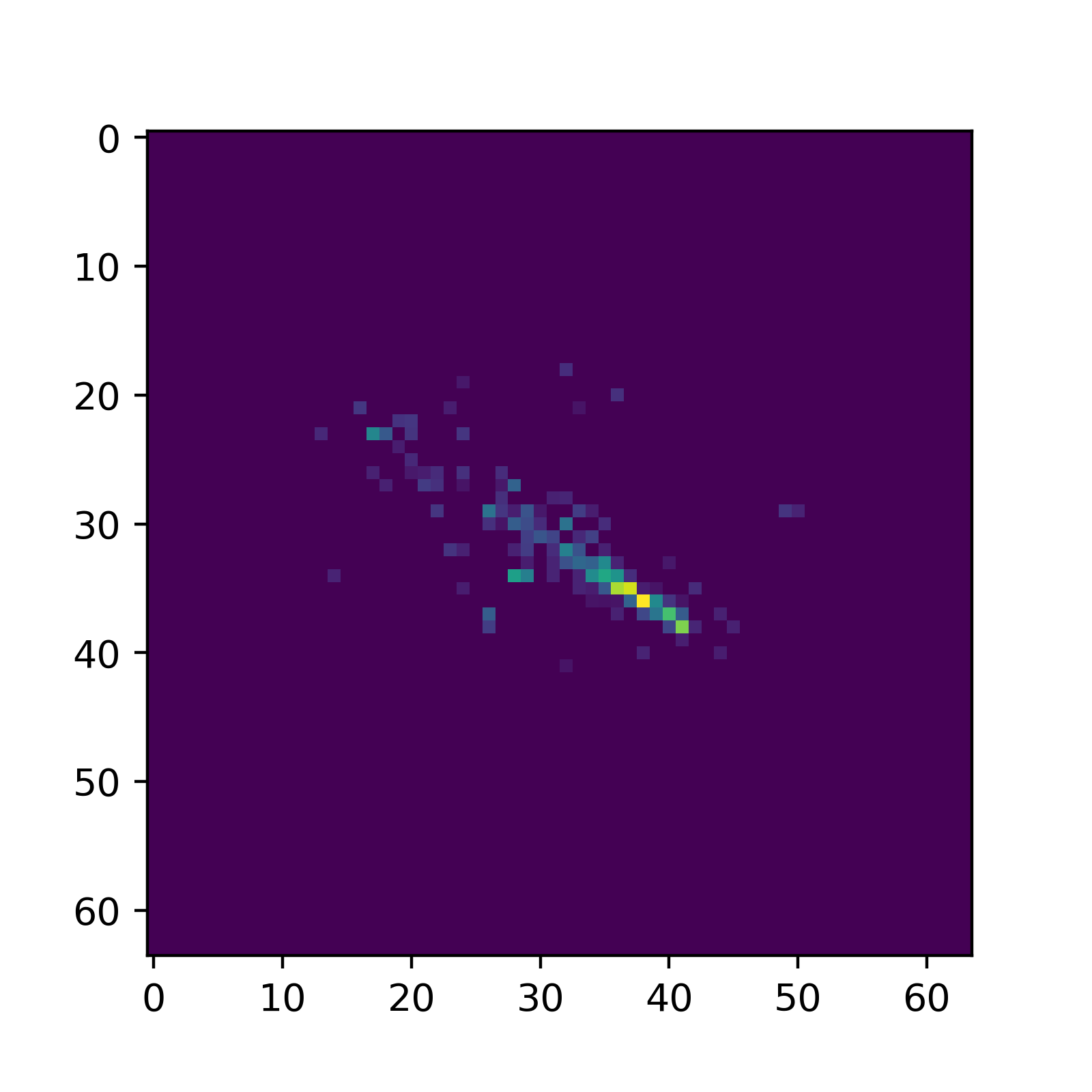}
}
\caption{Left: Imaging of $\gamma$-ray Cherenkov light in a telescope (3TeV); Right: Imaging of proton Cherenkov radiation in a telescope (9TeV)}
\label{Fig.7}
\end{figure}

The prevailing methodology for discriminating between high-energy gamma rays and background cosmic rays leverages the intrinsic differences between electromagnetic and hadronic showers. When photons impinge upon the Earth's atmosphere, they predominantly induce electromagnetic showers, which in contrast to the hadronic showers triggered by protons, yield more uniform detector images. Specifically, in the data recorded by our detectors, these images are represented as pixel arrays (as illustrated in Fig. \ref{Fig.7}). For a fitted ellipse, conventional methods compute the image moments of these pixels. If the shower is made up of photons or protons, it can be determined by comparing the major axis and minor axis of the fitted ellipse. Since photon-generated electromagnetic showers have a more concentrated energy distribution, their pixel layouts follow linear trajectories. Contrarily, hadronic showers brought on by protons exhibit more core hits, resulting in pixel patterns that are significantly spread and uneven. The main goal of current background suppression work is to distinguish between these two categories of image structures with accuracy. CNNs, in particular, have produced exceptional results in image identification using deep learning approaches. Consequently, compared to conventional curve-fitting techniques, they offer a more accurate discrimination capability.

\begin{figure}[htbp]
\centering
\subfigure[Modified ResNet-18 Model.]{
	\label{Fig.4.1}
	\includegraphics[width=0.95\textwidth]{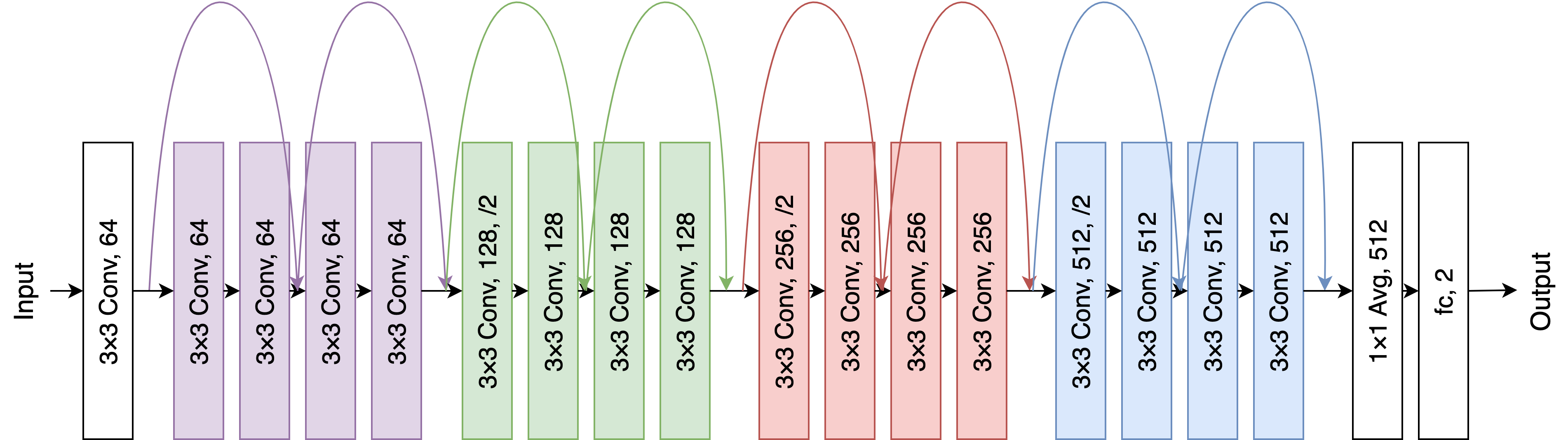}
}
\subfigure[Detailed Structure of Modified ResNet-18.]{
	\label{Fig.4.2}
	\includegraphics[width=0.7\textwidth]{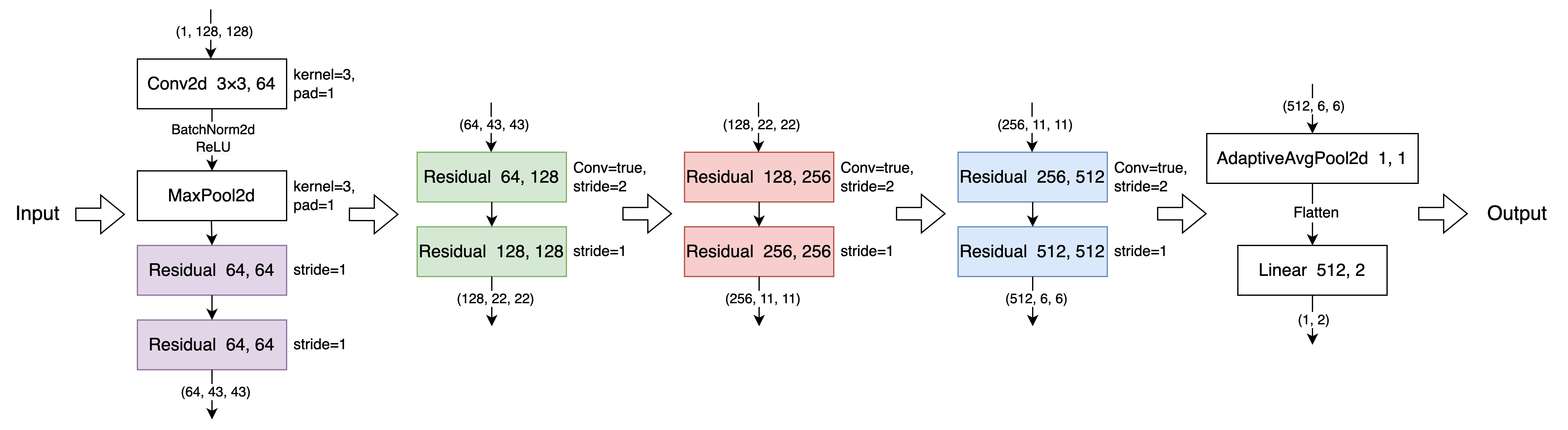}
}
\subfigure[Internal Structure of Residual Block.]{
	\label{Fig.4.3}
	\includegraphics[width=0.25\textwidth]{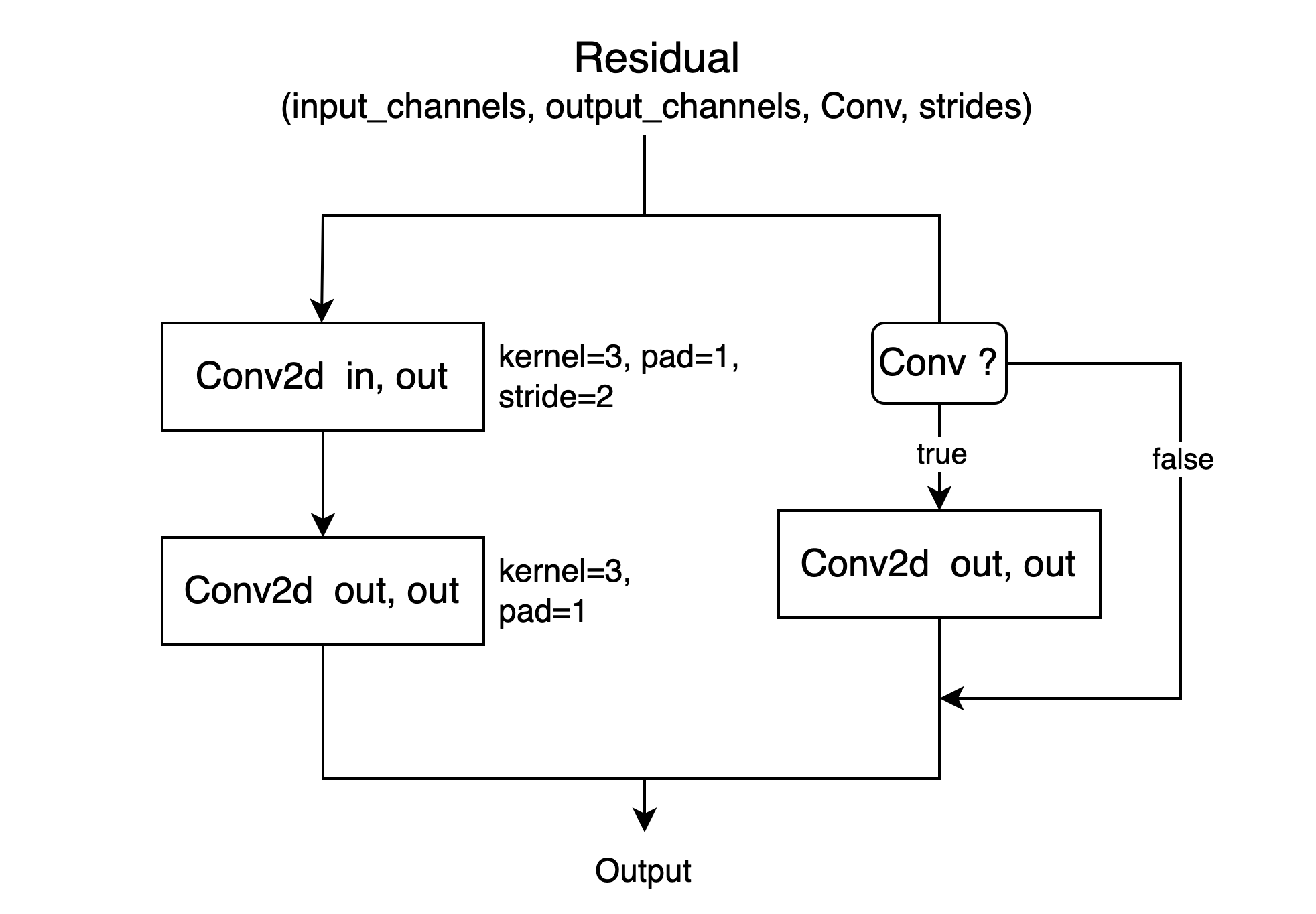}
}
\caption{The modified ResNet-18 Model used in the background suppression work, and we rename it to ParticleNet. Through four Residual blocks, ParticleNet increases the input data's initial dimensions to 512, and then the model processes it after a global pooling layer.}
\label{Fig.4}
\end{figure}

In the task of background suppression, we have adapted the ResNet-18 residual convolutional neural network model, as depicted in Fig. \ref{Fig.4}, to meet our specific objectives. The tailored architecture of our model is illustrated in Fig. \ref{Fig.4.1}. For a more comprehensive understanding of the Residual Block component, we also offer an expanded view in Fig. \ref{Fig.4.2} and Fig. \ref{Fig.4.3}. Unlike the canonical ResNet-18 model, our customized version employs smaller convolutional kernels with designated strides for the initial data convolution. This choice is motivated by the fact that the region of interest in our detector images is generally confined to a narrow central zone. The use of smaller kernels proves advantageous for capturing critical shape and edge features within this focal region, as substantiated by improved performance in subsequent evaluations.

The output from our model is formulated as a two-dimensional vector that quantifies the probabilities of a particle being categorized as either a signal (photon) or background noise (proton). This vector undergoes automatic normalization to ensure that its components sum to one, thus enhancing interpretability. The normalized probabilities $s_k$ are computed according to the equation:
\[
s_k = \frac{e^{z_k}}{\sum_{j=1}^{K} e^{z_j}}
\]
Here, $z_k$ denotes the raw output value corresponding to the $k^{th}$ class, and $K$ signifies the total number of classes. In our particular application, $K=2$ , representing the categories of signal and noise. This normalization ensures that the output probabilities are mutually exclusive and exhaustive, summing to 1, which aids in yielding more interpretable outcomes.

To evaluate the divergence between the model's predicted probability distribution and the actual probability distribution, we employ cross-entropy as our loss function, expressed as follows:
\[
CE(y, p) = -\frac{1}{N} \sum_{i=1}^N \sum_{j=1}^C y_{ij} \ln(p_{ij})
\]
Empirical results indicate that the cross-entropy loss function is highly sensitive to deviations in the model's predicted probabilities. Furthermore, it facilitates more rapid convergence during the training phase of the model.

\begin{figure}[htbp]
\centering
\subfigure[Photon image captured by a single water lens; Most of the PMTs (in purple) are not triggered, only a few PMTs within the white frame have captured signals.]{
	\label{Fig.6.1}
	\includegraphics[width=0.3\textwidth]{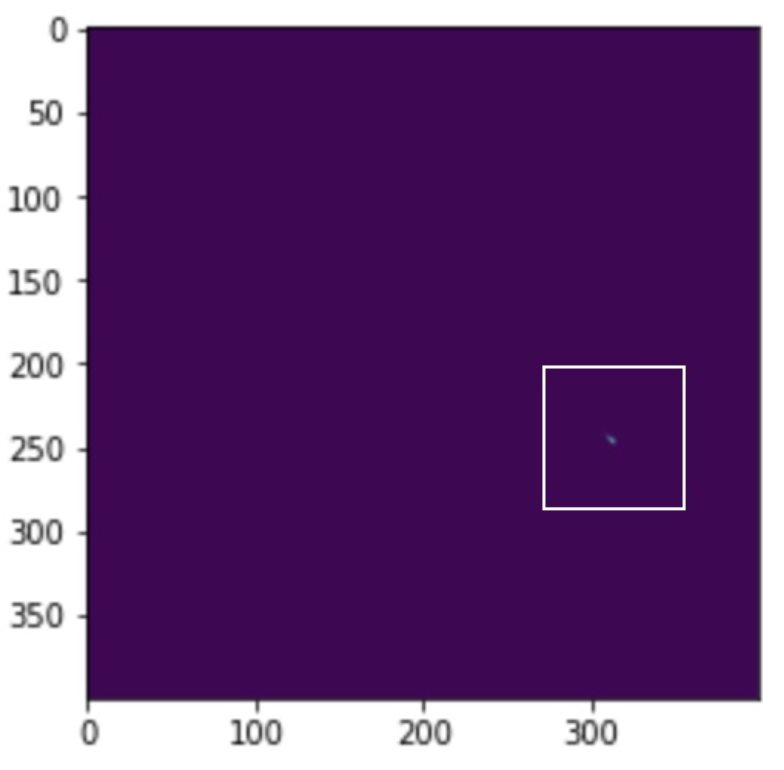}
}
\subfigure[Cropped pixel image centered on the location, which corresponds to the part enclosed by the white frame in Fig.\ref{Fig.6.1}.]{
	\label{Fig.6.2}
	\includegraphics[width=0.3\textwidth]{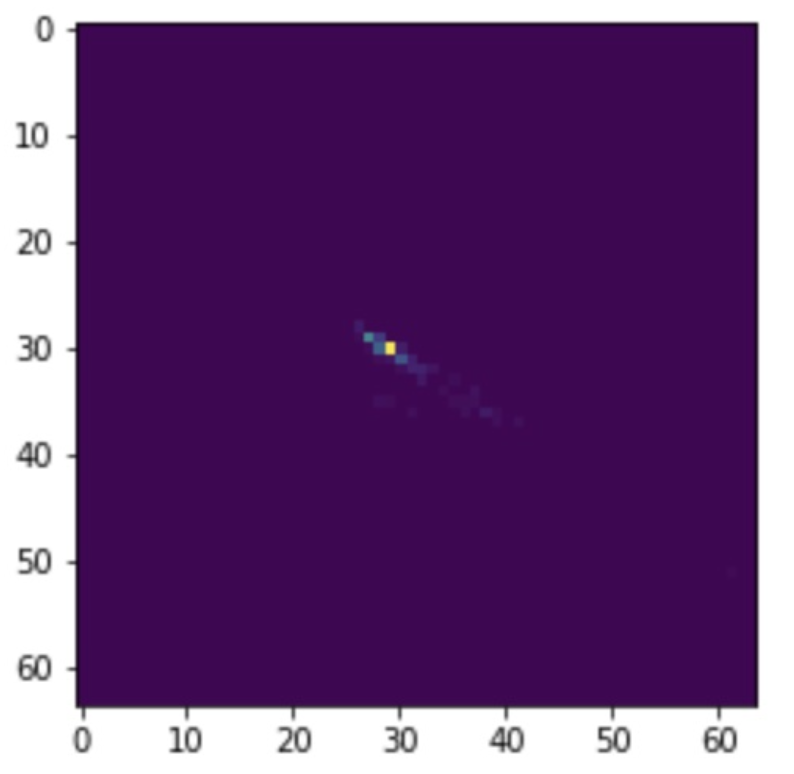}
}
\subfigure[Stitched version of four processed detector pixel images, with the processed image on the left situated in the white frame at the upper-left corner.]{
	\label{Fig.6.3}
	\includegraphics[width=0.3\textwidth]{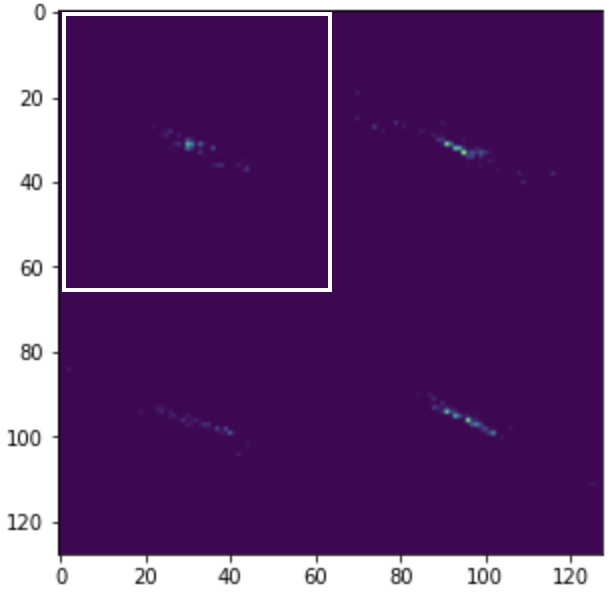}
}
\caption{Data preprocessing work involving the cropping and stitching of pixel photos.}
\label{Fig.6}
\end{figure}

The physical context was taken into account during the data preparation. Considering that electromagnetic cluster and hadron cluster will primarily create differences in the degree of image standardization, we combined the actual background of the current work: as a result, the absolute positions between different detectors become less important. For this reason, we decided to only retain a small part of the image (Fig.\ref{Fig.6.1}) centered on the actual signal area (Fig.\ref{Fig.6.2}), and stitch the remaining image together (Fig.\ref{Fig.6.3}). In the end, the model performed and processed the data above expectations accurately and precisely, providing a satisfying answer to this physical problem based just on intensity and relative positions.

\begin{figure}[htbp]
\centering
\includegraphics[width=0.7\linewidth]{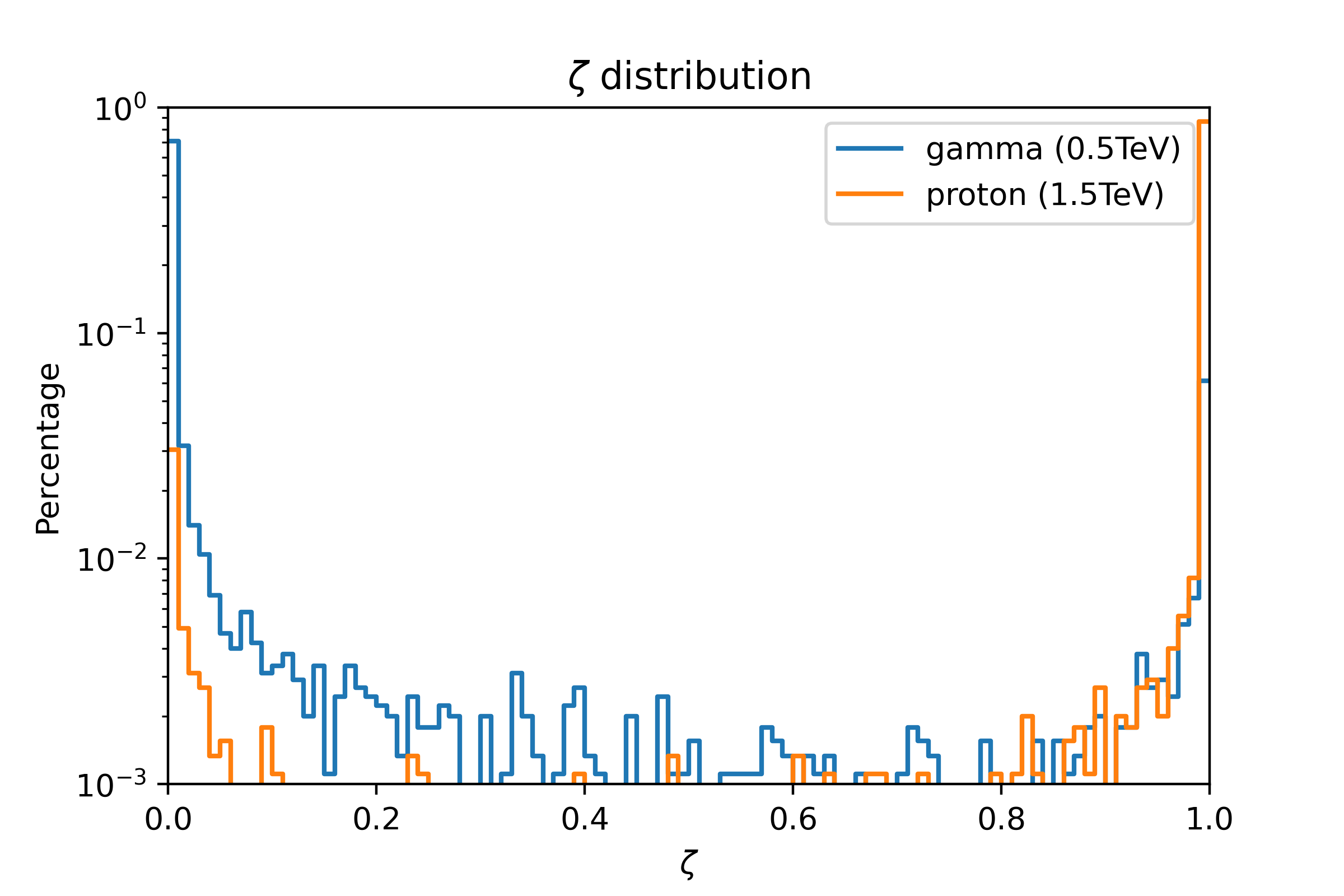}
\caption{Distribution of $\zeta$ values for photons and protons in simulated signals. The distribution of photons is concentrated around $\zeta = 0$, while the distribution of protons is centered around $\zeta = 1$.}
\label{Fig.zeta_distribution}
\end{figure}

To demonstrate the model's effectiveness and performance, we introduce a threshold factor, denoted by $\zeta_0$, and set its value to 0.5 as a criterion for interpreting the classification of incoming particles. In this context, $\zeta$ refers to the second component of the model's two-dimensional output vector. In an ideal scenario, the output for a photon signal should be aligned at $\zeta_{\gamma} = 0$, whereas the output for a proton signal should converge to $\zeta_{proton} = 1$. However, in practice, any intersection between the distribution curves of $\zeta_{\gamma}$ and $\zeta_{proton}$ would signify the model's error rate in classification, as illustrated in Fig.\ref{Fig.zeta_distribution}.
\end{multicols}

\begin{multicols}{2}
We retain the events identified as photons, while excluding those identified as protons. At the same time, we also record the correct identification rates for both photon and proton events. The quality factor $Q$ is defined as follows:
\[
Q=\frac{\varepsilon^{\gamma}_{s}}{\sqrt{\varepsilon^{p}_{s}}}
\]
Here, $\varepsilon^{\gamma}_{s}$ represents the survival rate of photons after discrimination, and $\varepsilon^{p}_{s}$ represents the survival rate of protons after discrimination. The higher the correct identification rate of the model, the larger the quality factor, indicating better background suppression.

\begin{figure}[htbp]
\centering
\begin{minipage}[t]{0.48\textwidth}
\centering
\includegraphics[width=\textwidth]{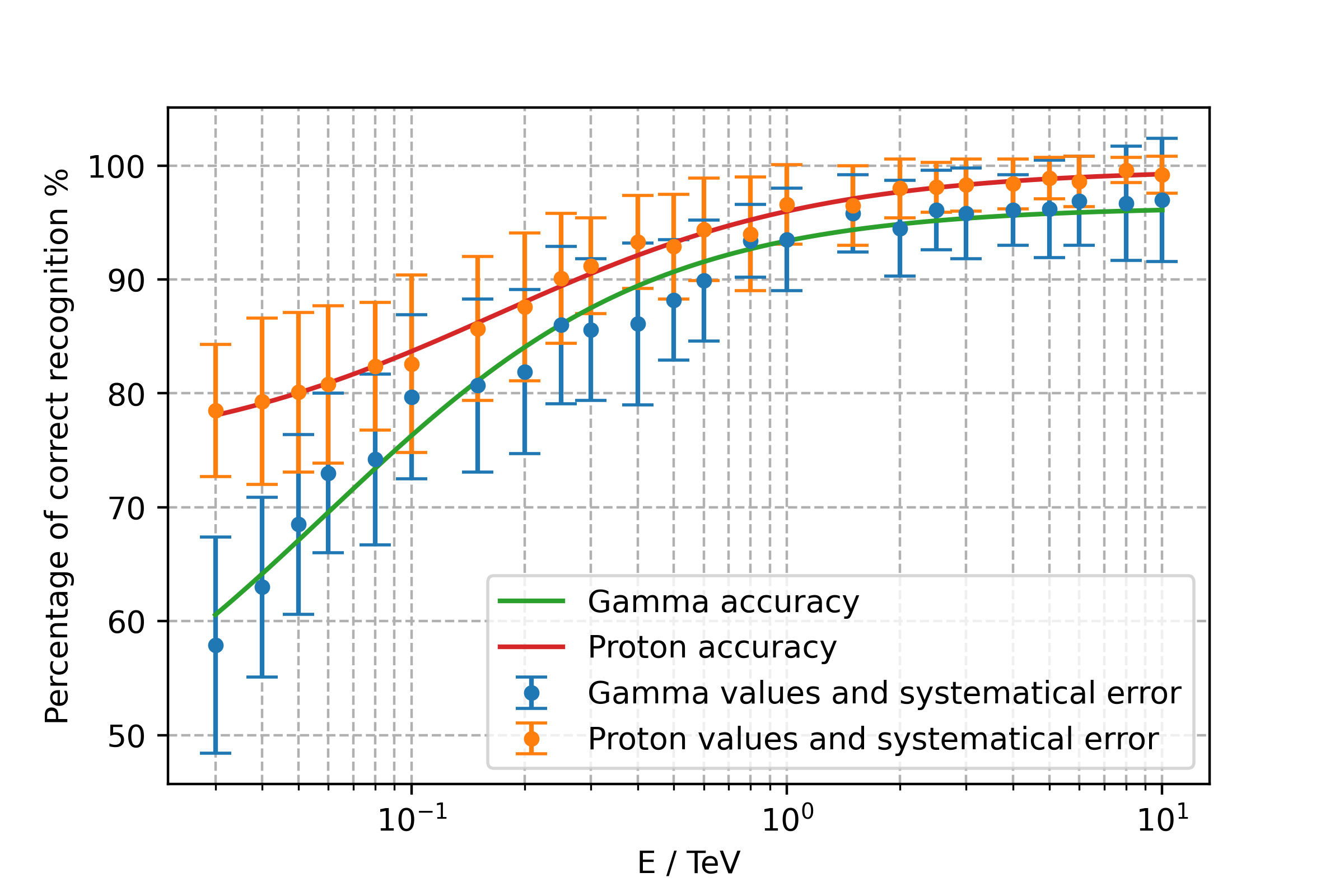}
\caption{The identification accuracy for photons/protons at different energy points in the test dataset. (The training process and results are from GPLearn, and variations may occur with different datasets.)}
\label{Fig.8}
\end{minipage}
\hfill
\begin{minipage}[t]{0.48\textwidth}
\centering
\includegraphics[width=\textwidth]{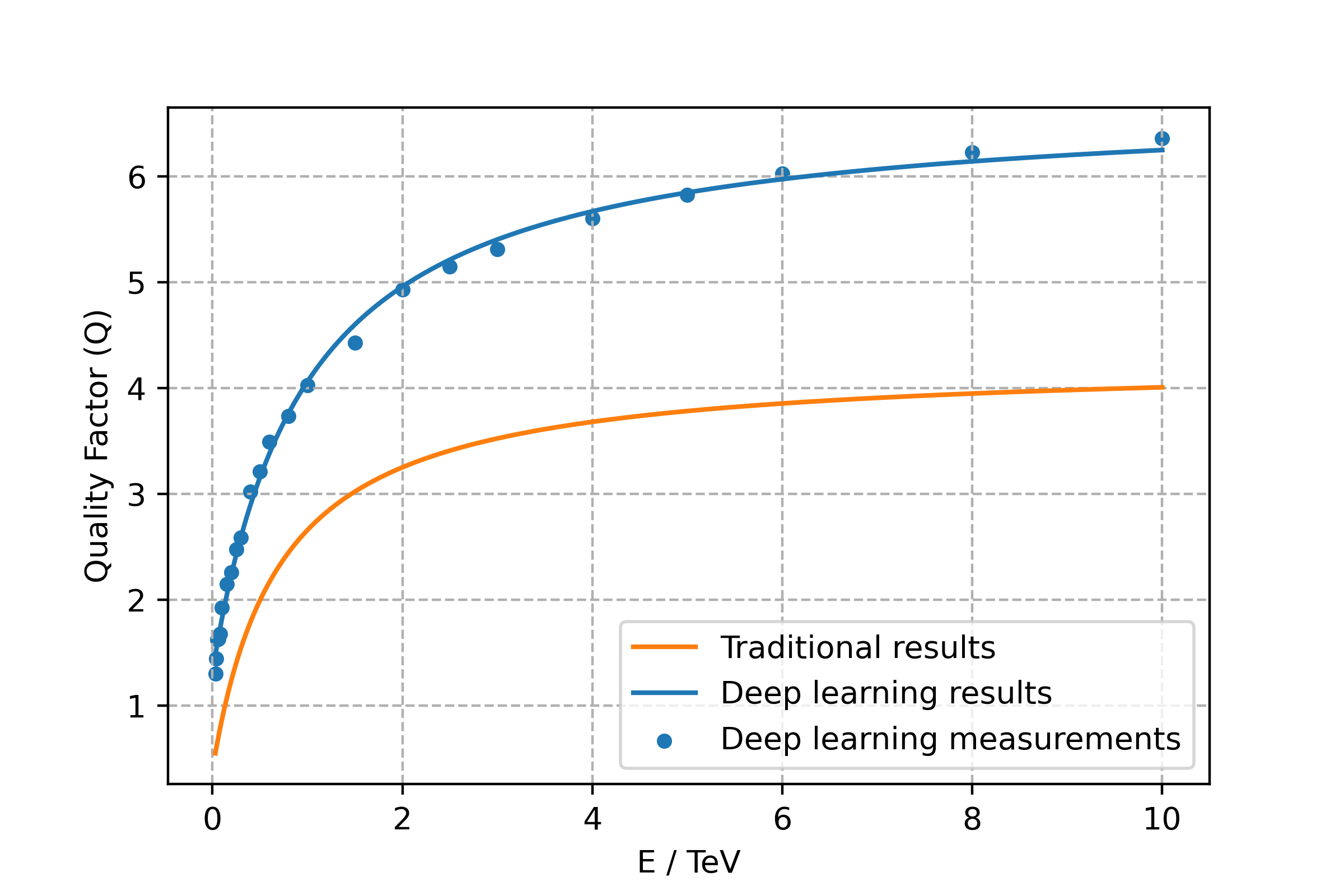}
\caption{Comparison of identification performance between deep learning and traditional methods for particle discrimination. (The deep learning results are provided by GPLearn; the traditional methods are calculated using the maximum ellipse fitting technique, which may vary slightly from current mainstream algorithms.)}
\label{Fig.9}
\end{minipage}
\end{figure}

Figure \ref{Fig.8} illustrates the model's accuracy in particle identification across varying energy thresholds. The yellow curve, corresponding to protons, demonstrates consistently high identification rates across the entire energy spectrum. Conversely, the blue curve, representing photons, exhibits a more variable identification rate, with a marked improvement at elevated energy levels. Existing research indicates that the threshold energy for the production of Cherenkov radiation is approximately 50 GeV. Near this energy threshold, even conventional electromagnetic showers are prone to generating anomalous results. As the energy level escalates, these showers exhibit increasing regularity, leading to a consequent enhancement in the model's identification accuracy.

The results from the test set indicate that the identification rates for both photons and protons exhibit a marked increase as particle energy rises. At lower energy levels, particles generate smaller atmospheric showers, resulting in a limited number of pixels captured by ground-based detectors. This scarcity of data hampers the model's ability to render accurate identifications. Conversely, as the particle energy escalates, the quality of images captured by the detectors improves substantially. This augmentation in data quality, coupled with an adequate number of parameters, leads to a significant uptick in the model's identification accuracy.

The yellow curve in Figure \ref{Fig.9} illustrates the performance gains achieved through the utilization of deep learning techniques, standing in stark contrast to the blue curve, which represents the efficacy of traditional methods. In the domain of low-energy particles, the quality of identification markedly ascends in tandem with rising energy levels, leading to a swift incline in the quality factor curve. Conversely, at higher energy thresholds, the model has ample raw parameters available for making accurate judgments, resulting in a plateauing of the result curve; it is worth noting that the model's identification rate has already approached a robust 98\% at this stage.
\end{multicols}

\begin{multicols}{2}
\section{Energy Reconstruction}
The model cannot function as intended if the dataset and model created in accordance with the background suppression work are utilized, and the related labels are changed to particle energy. The reconstructed particle energy is related to the incidence zenith angle, the distance from the cluster core, and the quantity of charge deposited in the detector after accounting for the pertinent physical background. Traditional approaches employ simulated data to determine the energy distribution function at each place and actual observational data to fit the following function \cite{RN96}:
\[
E_{erc}=f(q,R,\theta)
\]
Because the absolute positional information of the relative detector array is required to calculate the distance and direction information of the signal to the actual incident center, and because this information was removed during the dataset construction process for background suppression, data retaining only relative positional information cannot reconstruct the final energy.

\begin{figure}[htbp]
\centering
\includegraphics[width=0.95\textwidth]{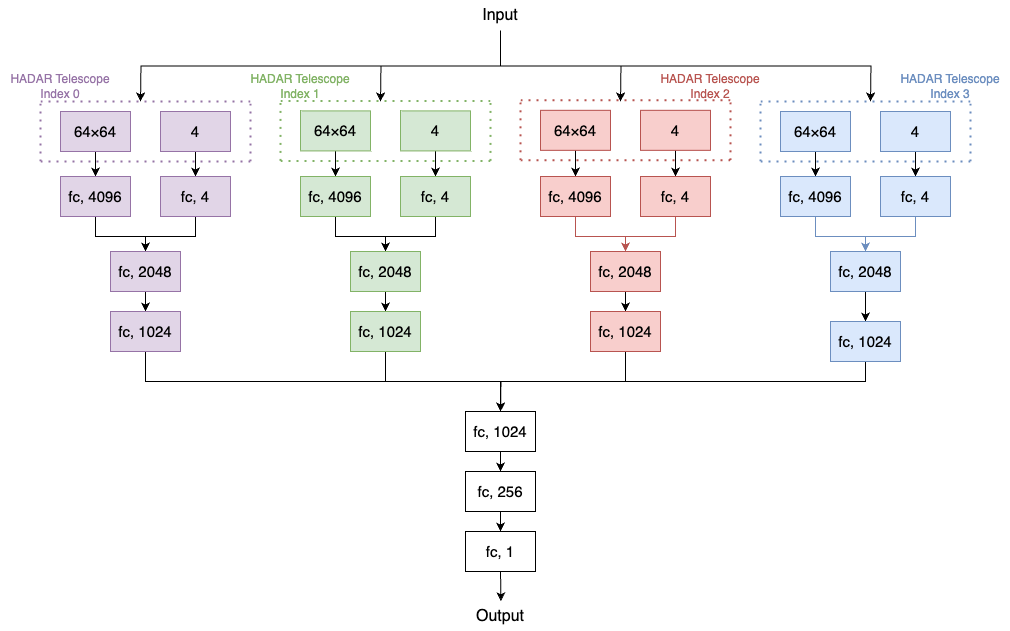}
\caption{The GoogLeNet model used in the energy reconstruction work. The model is further processed by combining the four detectors after processing the charge pictures and relative location data from four distinct detectors with fully linked layers independently.}
\label{Fig.GoogLeNetModel}	
\end{figure}

As a result, we added the relative position of the pixel centroid and the absolute position of the detector array as new inputs to the model, based on the data preprocessing work employed in background suppression work. The model was created as depicted in Figure \ref{Fig.GoogLeNetModel}, taking into account that energy resolution does not require particle classification work: four detector photos are processed independently, combined after a series of fully connected layers, and the final result is represented by a one-dimensional vector. This model, which does not include convolutional layers, not only retains a high level of reconstruction accuracy but also significantly speeds up calculation.

Expanding upon the data preprocessing techniques deployed in our background suppression task, we have incorporated additional variables: the relative positions of pixel centroids and the absolute coordinates of the detector array. This is in recognition of the fact that energy is fundamentally correlated with the amount of charge deposited. Accordingly, our model architecture, as illustrated in Figure \ref{Fig.GoogLeNetModel}, processes images from four individual detectors in parallel. These processed outputs are subsequently fused through a sequence of fully connected layers, culminating in a one-dimensional output vector.

\begin{figure}[htbp]
\centering
\begin{minipage}[t]{0.48\textwidth}
\includegraphics[width=\textwidth]{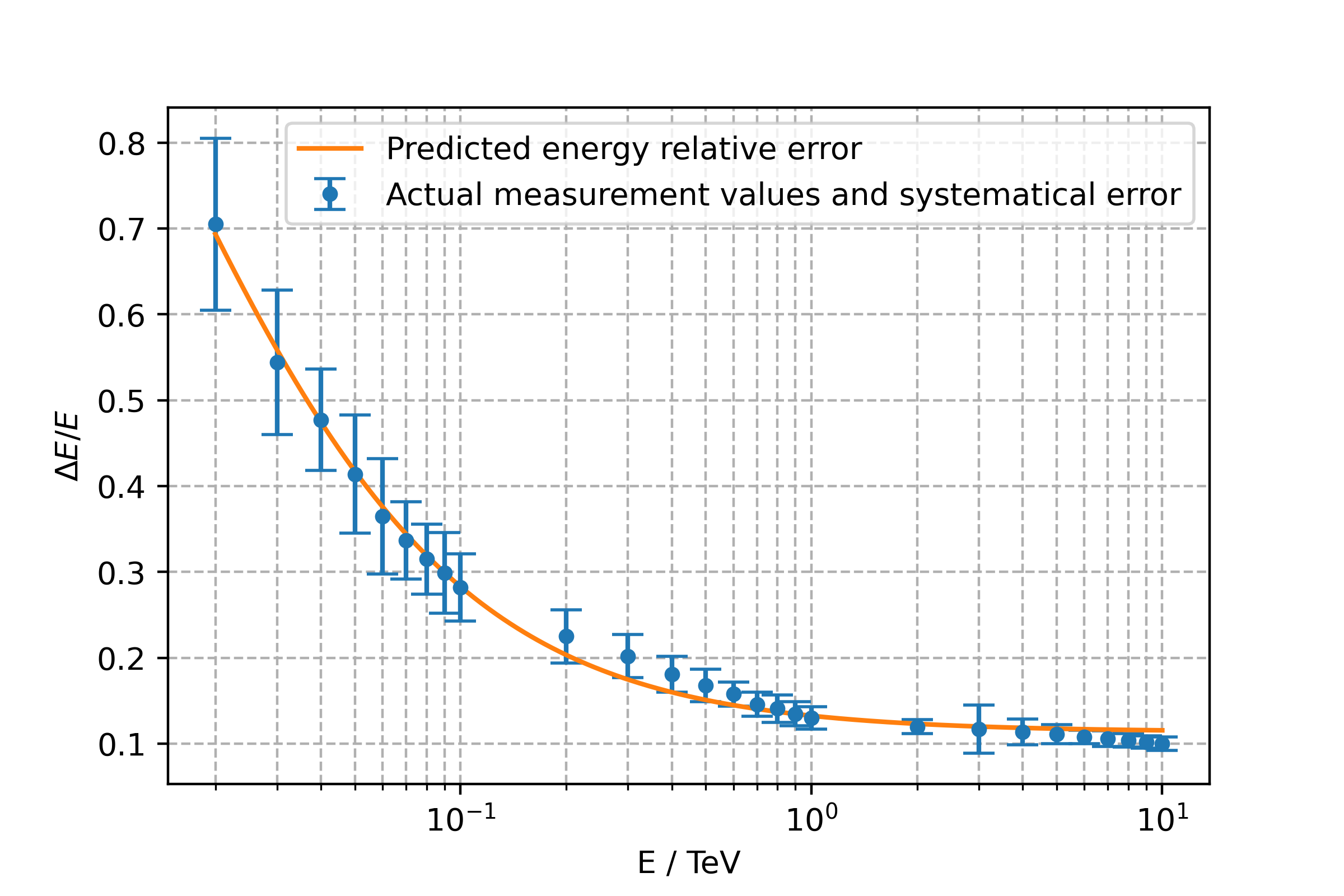}
\caption{Energy reconstruction's relative error distribution. The relative error rapidly reduces with particle energy until it reaches a value of 10.0\% at 10 TeV. By comparison, the relative error for the conventional technique is around 20\% at 1 TeV.}
\label{Fig.12}
\end{minipage}
\hfill
\begin{minipage}[t]{0.48\textwidth}
\includegraphics[width=\textwidth]{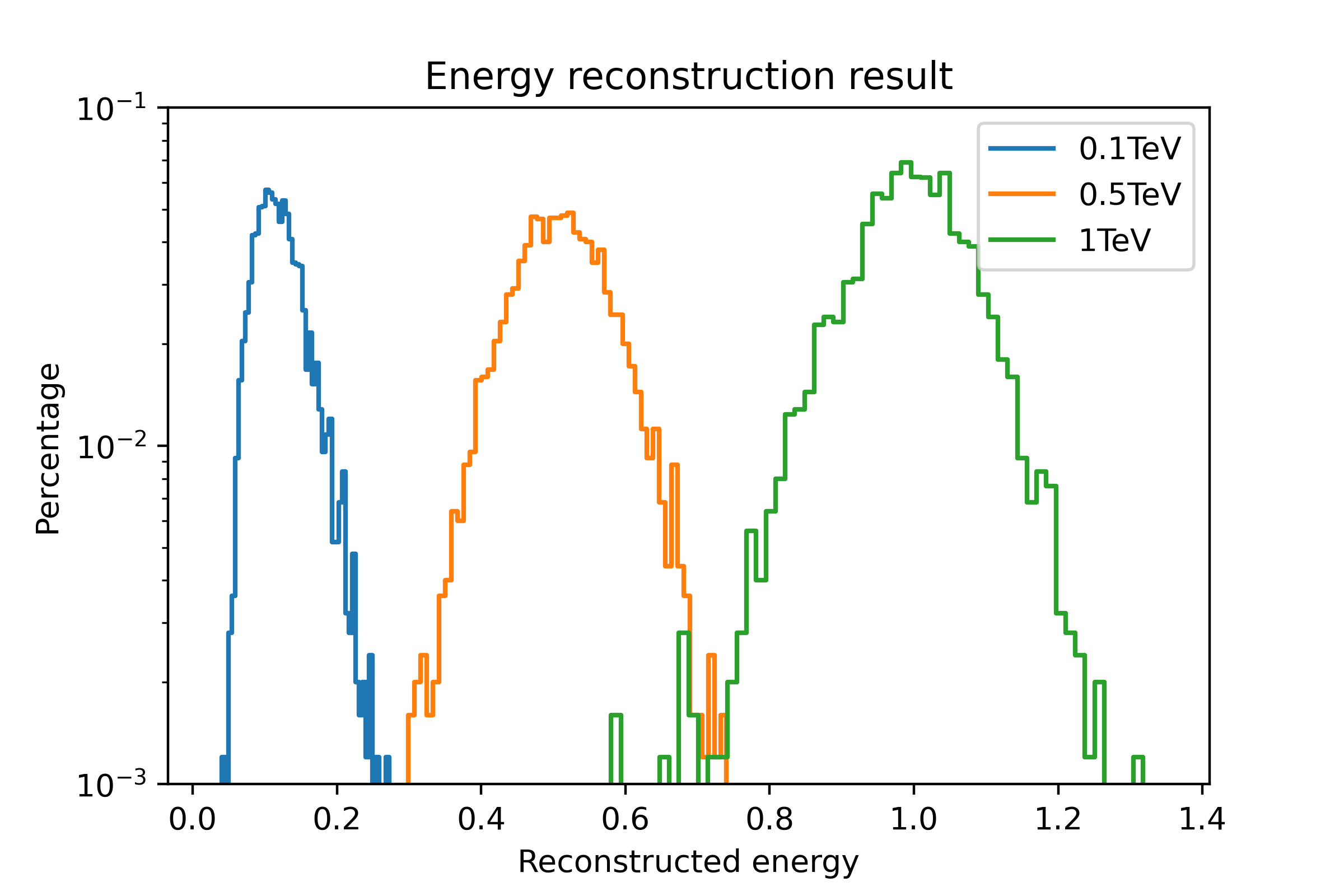}
\caption{Energy prediction results using deep learning methods. Different color curves barely overlap one another, showing that the model can almost entirely differentiate gamma rays at the appropriate energy levels.}
\label{Fig.13}
\end{minipage}
\end{figure}

In the test dataset, we employ both the model-predicted energy and the actual particle energy to compute the associated energy resolution, denoted as $\Delta E/E$. As illustrated in Figure \ref{Fig.12}, our findings reveal a trend of improving energy resolution with increasing particle energy.

To provide an intuitive assessment of our model's efficacy, we analyzed 2,500 samples from the test dataset, each representing a different energy level. We generated a distribution curve to illustrate the model's energy prediction performance. On the x-axis, we plotted the predicted energy values, while the y-axis represents the probability associated with each specific energy prediction. To facilitate easier interpretation, we normalized the data for each discrete energy level, ensuring that the integral of each curve across all energy points sums to unity.
 
Figure \ref{Fig.13} illustrates the distribution curves for energy prediction at 500 GeV intervals. Areas of overlap between these curves signify regions where the model's estimations may deviate from actual conditions, constituting a principal constraint on the accuracy of energy resolution. Importantly, our model achieves a relative error of just 13.0\% at an energy level of 1 TeV. Even more remarkably, when we ascend to higher energy levels—such as 10 TeV—the relative error narrows further to an impressive 10.0\%. This value approaches the theoretical limits of energy resolution achievable with Cherenkov telescopes, underscoring the significant strides our research has made.

\end{multicols}

\begin{multicols}{2}
\section{Direction Reconstruction}
After atmospheric Cherenkov radiation reaches ground-level detectors, it is refracted through our specialized water lenses before being captured on the image plane. Calculating the path of this refracted light is complex. Our initial effort focuses on estimating the shower core's projected position on this image plane. By calculating the fitting ellipse of the four detector images and identifying the location where clustering occurs based on the weighted intersection of the long axes of each ellipse, conventional methods (such as Hillas reconstruction \cite{RN83}) can determine where clustering occurs and ultimately the direction of incidence of $\gamma$-rays \cite{RN84} \cite{RN87}.

Unlike the energy reconstruction task, both the intensity and absolute position of each pixel are intricately linked, serving as indirect indicators of the gamma rays' incident direction. 

However, it is impractical to fully create a pixel matrix based on actual positions because each pixel corresponds to a photomultiplier tube with a 1.25 cm diameter and a 100 m separation between detectors, making it impossible to create a two-dimensional matrix with such a large input size. We continue to use the cropping method in response to this. In contrast to the background suppression work, each pixel after cropping is thought of as a three-dimensional data set because it contains the deposited charge corresponding to that pixel as well as its absolute coordinates prior to cropping, which have been converted using the precise coordinates of the experimental site. Our tests demonstrate that the model's performance is much enhanced when the absolute positioning information of pixels is retained.

\begin{figure}[htbp]
\centering
\includegraphics[width=0.95\textwidth]{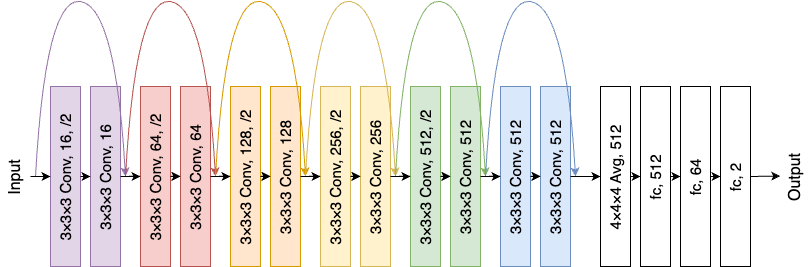}
\caption{Model used for direction reconstruction: AngleNet Model. Through six convolutions, AngleNet increases the input's original dimension to 512, and then processes it through a fully connected layer after a global pooling layer. AngleNet makes use of Residual Blocks that are appropriate for three dimensions, just like ParticleNet does. This helps to some extent with the under-fitting issue with the model and considerably improves its ability to forecast the future.}
\label{Fig.5}	
\end{figure}

To accommodate the specialized input dataset, we employ a CNN model featuring three-dimensional convolutional layers, as illustrated in Figure \ref{Fig.5}. The raw data undergoes a sequence of six up-sampling convolutional layers, each followed by a pooling layer. This processed data is then fed into a linear regression layer, culminating in a two-dimensional vector output. After normalization, this vector serves as an azimuthal direction vector in the Earth's plane. The model subsequently computes the corresponding incident angle based on this vector.

The error angle distance $\Omega$ is calculated from the azimuthal angle through spherical coordinate transformation.
\[
cos\Omega=sin\theta sin\phi \cdot sin\theta sin\phi ' + sin\theta cos\phi \cdot sin\theta cos\phi ' + cos^2\theta
\]
Here, $\phi$ represents the true incident azimuthal angle, while $\phi'$ represents the azimuthal angle reconstructed by the model.

\begin{figure}[htbp]
\centering
\includegraphics[width=0.7\linewidth]{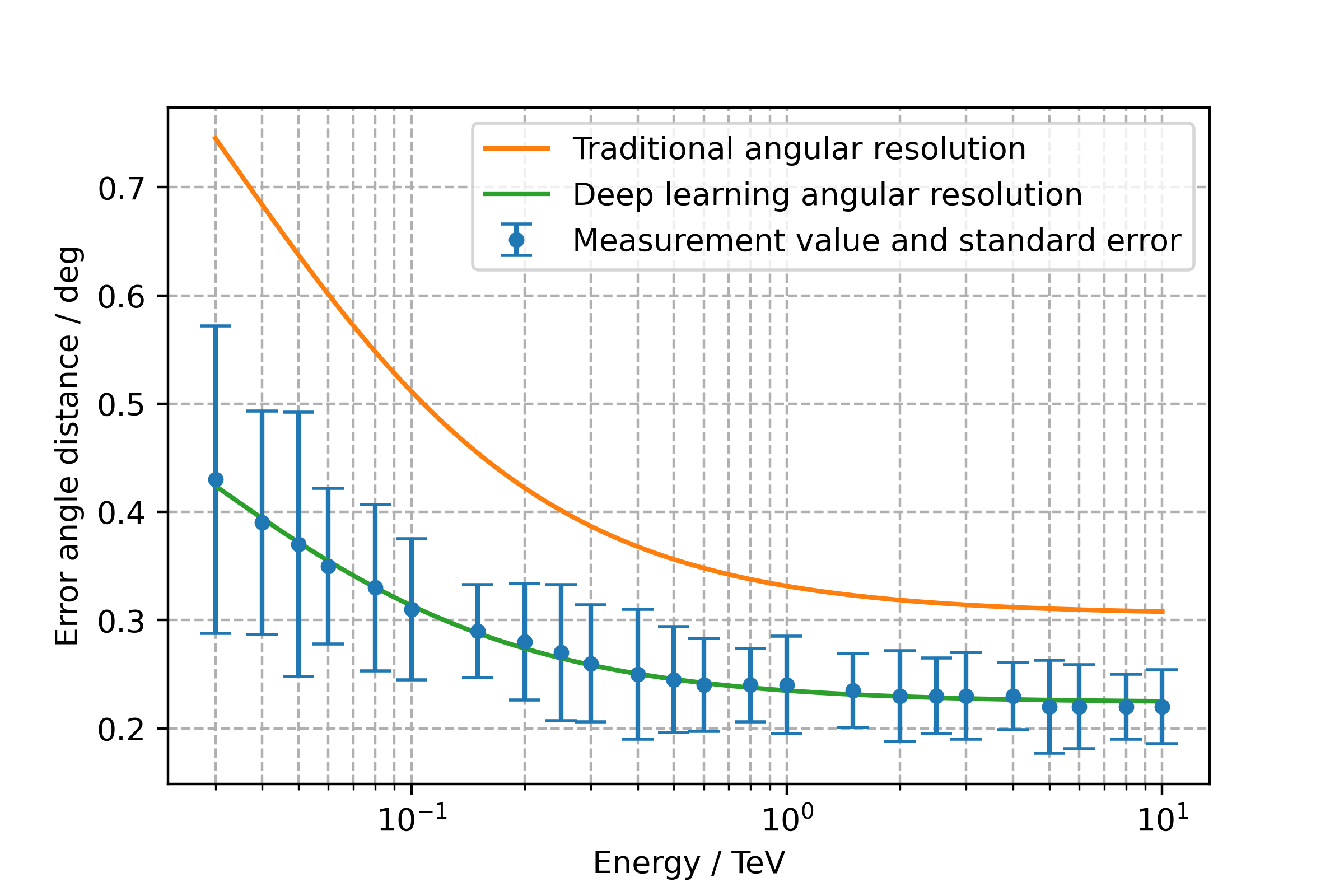}
\caption{Angular distance difference between reconstructed and true incident directions at different energies, in degrees. Deep learning techniques have a larger potential for the outcomes of angle reconstruction than conventional elliptical fitting techniques since they rebuild this physical process by looking at relationships among nearly all of the original data. (Deep learning results are provided by GPLearn; traditional method data is sourced from the literature \cite{RN65})}
\label{Fig.16}
\end{figure}

All our simulations are based on an incident zenith angle of $\theta = 20^\circ$ . The results, plotted in Figure \ref{Fig.16}, reveal that the deep learning method we employed yields approximately a 60\% improvement in accuracy over traditional methods as cited in \cite{RN65}. Notably, the reconstruction error decreases substantially with increasing energy levels. This enhanced accuracy is attributable to the greater number of photons collected by the telescope at higher energy levels, which in turn allows for a more precise reconstruction of the event direction. In the energy range above 1 TeV, our model has already reduced the angular error to below 0.2 degrees, providing a robust foundation for subsequent research endeavors.
\end{multicols}

\begin{multicols}{2}
\section{Simulated measurements of the Crab Nebula}
\subsection{Sensitivity measurements}
The Crab Nebula serves as a prototypical supernova remnant, emitting across a wide range of wavelengths including radio, X-ray, and gamma-ray bands. The study of this celestial object is pivotal for gaining insights into the physical mechanisms underlying supernova explosions, the generation and acceleration of cosmic rays, and addressing important scientific queries such as the nature of dark matter. Among the numerous physical attributes associated with the Crab Nebula that demand precise measurement, sensitivity stands out as a crucial parameter. The level of sensitivity directly impacts the quality—both in terms of accuracy and precision—of our estimates concerning various physical phenomena related to the Crab Nebula.

In preceding sections, we delved into the deployment of deep learning methodologies for the sophisticated analysis of detector data, achieving a marked improvement in accuracy. In this subsequent segment, our focus shifts to the specific application of deep learning in enhancing the sensitivity measurements related to the Crab Nebula. We will illustrate how these advanced computational techniques outperform traditional methods, offering a more precise and reliable assessment of this crucial parameter. Additionally, we will juxtapose our findings with those obtained from other Cherenkov imaging telescope experiments to contextualize the efficacy and innovation of our approach.

In order to calculate sensitivity, we conducted a simulation that accounts for both gamma rays and scattered protons across an energy spectrum ranging from 100 GeV to 10 TeV. The simulated sampling area is set at $S_{sample}=800\times 800m^2$, and we used a fixed zenith angle of 20 degrees. The azimuthal angle in the simulation varies between 0 and 180 degrees. Additionally, we integrated geographical factors specific to the Yangbajing site and the celestial trajectory of the Crab Nebula. Based on these considerations, we estimated the annual observation time dedicated to the Crab Nebula to be approximately 320 hours, which translates to $T_{crab,obs}=1.152 \times 10^6s$.

For gamma rays, we can calculate the effective number of events coming from the Crab Nebula received by the HADAR experiment in one year at energy $i$ using the equation
\[
N_\gamma^{1yr}[i]=T_{crab,obs} \cdot S_{sample} \cdot F_{crab} \cdot \frac{N_{\gamma,sim}[i]}{N_{\gamma,sim,all}[i]} \cdot \varepsilon_\gamma
\]
Here, $F_{crab}=\int_i^\infty 2.83\cdot 10^{-11}phcm^{-2}s^{-1}TeV^{-1} \cdot (E/TeV)^{-2.62} dE$represents the integrated flux of the Crab Nebula for energies greater than $i$. $N_{\gamma,sim}[i]$ is the number of effective simulated photon events in energy range $i$, $N_{\gamma,sim,all}[i]$ is the total number of simulated photon events, and $\varepsilon_\gamma$ represents the ratio of events located within the angular resolution, set at 68\%.

For scattered protons, the effective number of observed events for one year can be calculated as
\[
N_{CR}^{1yr}[i]=T_{crab,obs} \cdot S_{sample} \cdot F_{CR} \cdot \frac{N_{CR,sim}[i]}{N_{CR,sim,all}[i]} \cdot \Omega_i
\]
Here, $F_{CR}$ represents the integrated flux of protons in cosmic rays. $N_{CR,sim}[i]$ is the number of effective simulated proton events in the energy range $i$, $N_{CR,sim,all}[i]$ is the total number of simulated proton events, and $\Omega_i$ represents the solid angle within the range of photon angular resolution.

\begin{figure}[htbp]
\centering
\includegraphics[width=0.75\linewidth]{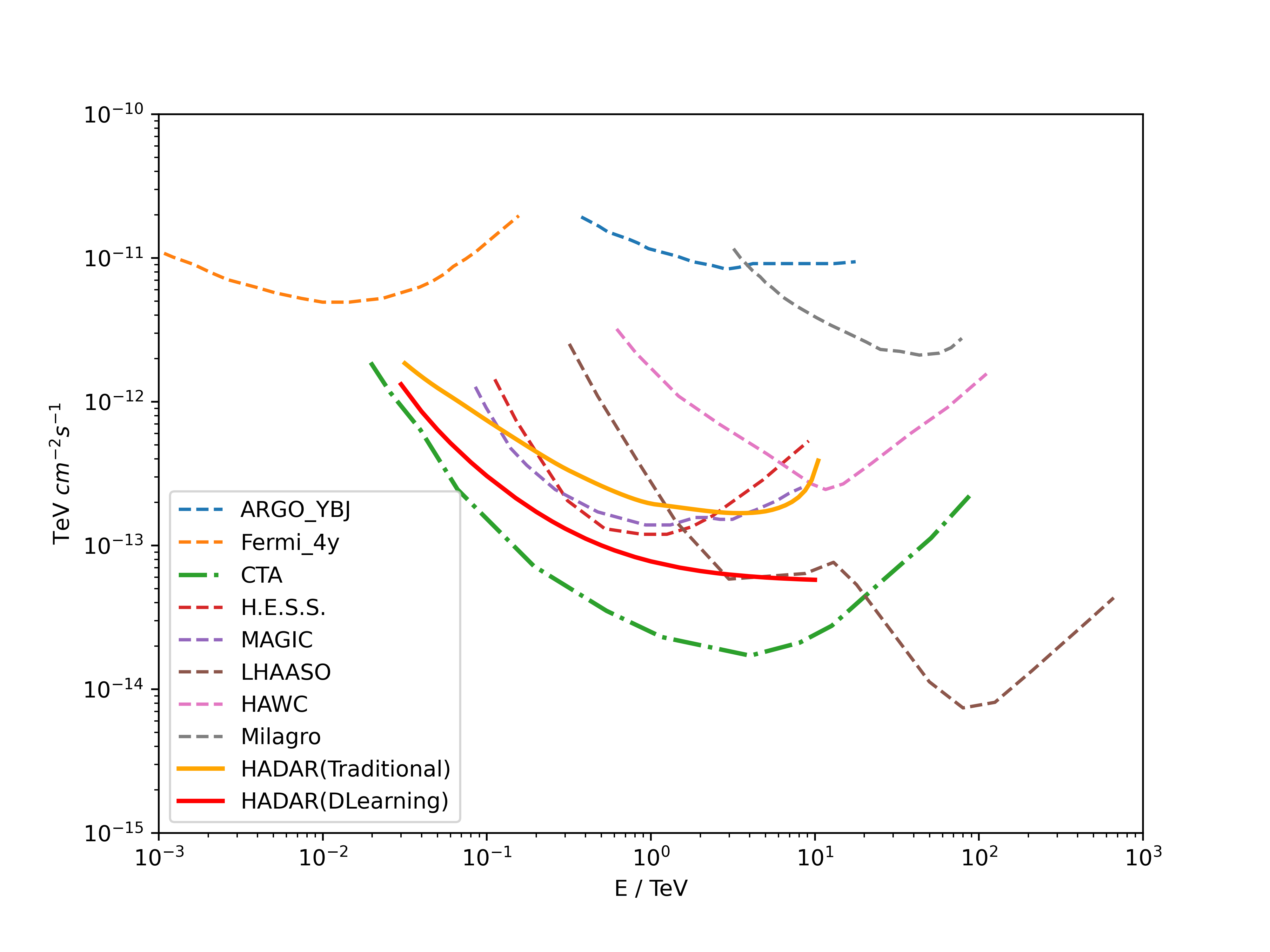}
\caption{Comparison of Sensitivity Between HADAR Experiment and Other Experiments. The HADAR experiment's sensitivity curve for the Crab Nebula with a 5-fold significance over a year (320 hours of useful observation time) is shown in the figure. The Crab Nebula sensitivity curves from experiments like the Fermi Satellite (one year of effective observation time); MAGIC, H.E.S.S. (50 hours of effective observation time); and ARGO-YBJ, HAWC (one year of effective observation time) are also shown in the figure for comparison.}
\label{Fig.17}
\end{figure}

Subsequent to the implementation of background suppression techniques and the exclusion of proton-like events, we can derive the integrated significance across the designated energy spectrum via Equation \cite{RN55}:
\[
S[i]=\frac{N_\gamma^{1yr}[i]}{\sqrt{N_{CR}^{1yr}[i]}} \cdot Q
\]
Utilizing the framework of Gaussian statistics, the sensitivity is defined as the minimal flux from the Crab Nebula required for the detector to register a signal at a 5-sigma significance level. Consequently, the specific sensitivity pertaining to the Crab Nebula for the HADAR experiment can be calculated using the formula:
\[
F_{sensitivity}[i]=F_{crab} \cdot \frac{5}{S[i]} \cdot i
\]
To offer a comprehensive perspective, we graphically delineate the calculated sensitivity metrics, both prior to and following the incorporation of deep learning algorithms. These are juxtaposed with corresponding data from other pertinent experiments \cite{RN56}\cite{RN58}\cite{RN59}, as depicted in Figure \ref{Fig.17}.

Following the integration of advanced deep-learning algorithms, the HADAR experiment has experienced a substantial enhancement in sensitivity within the low-energy domain. Although it has yet to attain the level of sensitivity exhibited by the CTA, the revamped HADAR experiment now stands as a formidable contender in juxtaposition with established initiatives such as MAGIC and H.E.S.S. Importantly, HADAR boasts a unique attribute: an expansive field of view not typically afforded by conventional reflective Cherenkov telescopes. This distinct advantage facilitates the real-time capture of transient celestial sources within its observational purview, rendering the outcomes particularly compelling.
\end{multicols}

\begin{figure}[htbp]
\centering
\includegraphics[width=0.7\linewidth]{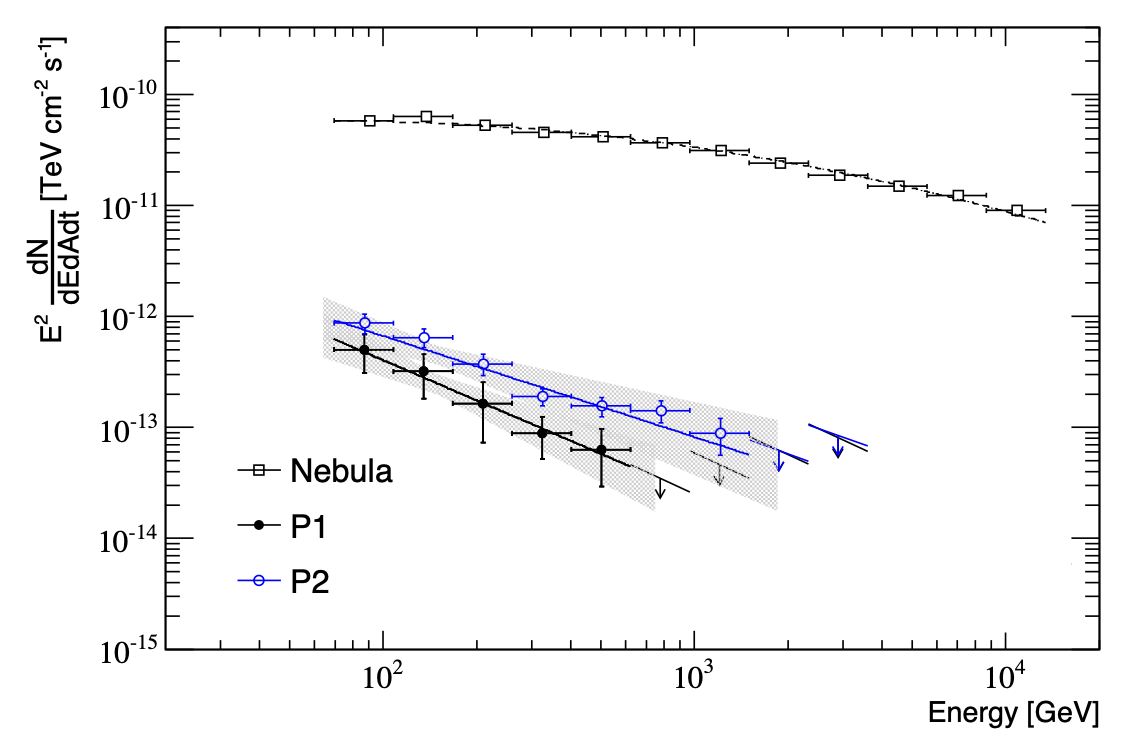}
\caption{The energy spectra for the Crab Nebula's main pulsar P1 (represented by black circles) and the interpulsar P2 (represented by blue circles) in the 70GeV-1.5TeV energy range have been measured by MAGIC. The energy spectrum of the Crab Nebula itself (depicted by hollow squares) is also shown for comparison. (Data sourced from \cite{RN54}).}
\label{Fig.18}
\end{figure}

\begin{multicols}{2}
\subsection{Observation of the Pulsar in the Crab Nebula}
The primary objective of this section is to undertake a rigorous analysis of the observational data pertaining to the Crab Nebula pulsar in the very-high-energy regime, as acquired through the MAGIC telescope \cite{RN60}. The intent behind calculating the significance of observations from the HADAR experiment lies in laying the groundwork for future inquiries into pulsar emissions at very-high-energy wavelengths.

We determined the spectral properties of the primary pulsar (P1) and interpulsar (P2) of the Crab Nebula using data obtained from the MAGIC telescope \cite{RN54}, as shown graphically in Figure \ref{Fig.18}. The formula for the integral flux for each energy band is as follows:
\[
F_{P1}=1.1\times10^{-11} \cdot (\frac{E/TeV}{0.15})^{-3.2} cm^{-2}s^{-1}
\]
\[
F_{P2}=2\times10^{-11} \cdot (\frac{E/TeV}{0.15})^{-2.9} cm^{-2}s^{-1}
\]

Following the implementation of background filtering procedures, only events characterized by gamma-like profiles are retained for subsequent analyses concerning the pulse signals of the Crab Nebula's primary pulsar (P1) and interpulsar (P2). To quantify the observational significance $S$ for the HADAR experiment, we employ the following equation:
\[
S=\frac{N_{on}-\alpha N_{off}}{\sqrt{\alpha(N_{on}+N_{off})}}
\]
Here, $N_{on}$ signifies the photon count emanating from the pulsar, while $N_{off}$ represents the background photon count. Remarkably, at the 100 GeV energy threshold, the dominant background source in observations of the Crab Nebula pulsar transitions from protons to gamma rays emanating from the Crab Nebula itself. Consequently, the term $N_{off}$ encapsulates both proton and non-pulsar gamma-ray backgrounds. The observation time ratio $\alpha$ is defined as $\alpha={t_{on}}/{t_{off}}$, where $t_{on}$, where $t_{on}$ and $t_{off}$ denote the durations of HADAR detector observations for the source and background, respectively.

Table \ref{Tab.MAGIC} and Table \ref{Tab.HADAR} delineate the observational datasets acquired from the MAGIC and HADAR experiments, respectively. These tables encapsulate the derived significances corresponding to distinct energy bands for both the main pulsar (P1) and the interpulsar (P2) of the Crab Nebula, as corroborated by references \cite{RN61} and \cite{RN62}.

Capitalizing on the robust data processing capabilities afforded by deep learning methodologies, the HADAR experiment demonstrates a notably superior level of significance within the low-energy domain as compared to the MAGIC experiment. Nevertheless, as the energy spectrum ascends, certain constraints, including equipment specifications and geographical considerations, result in a discernible performance disparity in comparison to MAGIC. Moving forward, we aim to augment the sensitivity of our apparatus through a multi-faceted approach, encompassing the expansion of the detector's effective observational area as well as the refinement of computational models to enhance angular resolution.
\end{multicols}

\begin{table}[htbp]
\centering
\caption{Five years of observational (320 hours of effective observation time) data from the MAGIC experiment shows corresponding significances for different energy ranges for P1 and P2. (Data sourced from \cite{RN54}).}
\label{Tab.MAGIC}
\begin{tabular}{c|c|c|c|c}
\hline
\hline
energy range & \multicolumn{2}{c|}{P1} & \multicolumn{2}{c}{P2} \\
\hline
[$GeV$] & $N_{ex}$ & Significance & $N_{ex}$ & Significance \\
\hline
100$-$400 & $1252\pm442$ & $2.8\sigma$ & $2537\pm454$ & $5.6\sigma$ \\
\hline
$>$400 & $188\pm88$ & $2.2\sigma$ & $544\pm92$ & $6.0\sigma$ \\
\hline
$>$680 & $130\pm66$ & $2.0\sigma$ & $293\pm69$ & $4.3\sigma$ \\
\hline
$>$950 & $119\pm54$ & $2.2\sigma$ & $190\pm56$ & $3.5\sigma$ \\
\hline
\hline
\end{tabular}
\end{table}
	
\begin{table}[htbp]
\centering
\caption{Deep learning algorithms are used to determine the comparable significance in the same energy range estimated from the observation results of P1 and P2 pulsars in the HADAR experiment over a year (320 hours of effective observation time).}
\label{Tab.HADAR}
\begin{tabular}{c|c|c|c|c}
\hline
\hline
energy range & \multicolumn{2}{c|}{P1} & \multicolumn{2}{c}{P2} \\
\hline
[$GeV$] & $N_{ex}$ & Significance & $N_{ex}$ & Significance \\
\hline
100$-$400 & $3698$ & $4.65\sigma$ & $6943$ & $7.31\sigma$ \\
\hline
$>$400 & $500$ & $0.708\sigma$ & $1432$ & $2.046\sigma$ \\
\hline
$>$680 & $175$ & $0.390\sigma$ & $584$ & $1.306\sigma$ \\
\hline
$>$950 & $87$ & $0.261\sigma$ & $321$ & $0.963\sigma$ \\
\hline
\hline
\end{tabular}
\end{table}

\begin{multicols}{2}
\section{Outlook}
With the emergence of large-scale AI language models, artificial intelligence has been progressively incorporated into a myriad of industrial applications. Within the realm of physics, data analytics stands as a burgeoning arena for the expansion of AI technologies. Leveraging its computational prowess, AI has the potential to uncover latent relationships within complex data sets, accelerate data-fitting procedures, and simulate realistic models for experimental measurements.

In high-energy physics, the synergy between AI and big data allows for the precise reconstruction of energy, momentum, and mass metrics for particles, a critical step towards elucidating the underlying properties and behaviors of subatomic particles. Additionally, exact spatial reconstructions are indispensable for tracking the decay products of heavy particles and for capturing rare events that could signal novel particles or interaction mechanisms. In the field of astronomy, AI technologies facilitate the real-time processing of astronomical signals, enabling more efficient studies of distant cosmic events and contributing to our understanding of the universe's origins and composition.

In this study, we effectively address complicated physical issues by integrating real-world physical contexts, using appropriate data preparation, and using appropriate models. Additionally, the excellent results from deep learning also serve to confirm the accuracy of the corresponding physical theories: the training process, which was built in accordance with real-world physical theories, both theoretically supports and verifies the deep learning results.

In the actual research process, we have demonstrated that:
\begin{enumerate}
\item The relative position and shape information of Cherenkov radiation can be used to discriminate particle types; at the same time, information contained in data that is not sensitive to absolute position will not be lost after slicing and cropping processes.
\item The energy information of the Cherenkov radiation reaction is not sensitive to the relative position information of pixels; moreover, energy can be approximately represented by a linear function of charge amount, incidence angle, and relative center distance.
\item The directional information of Cherenkov radiation is sensitive to the absolute position of pixels; therefore, retaining the absolute position information of pixels is crucial for correctly inferring the initial direction of particles when processing image data.
\end{enumerate}

Deep learning techniques have simplified and improved some labor-intensive physical reconstruction procedures. But because physical data features are typically sensitive to a limited scale range, adopting small-scale convolutional kernels aids the neural network in more acutely recognizing features. The network can more efficiently gather data on global feature information by expanding its dimensions. The cropping of data is made possible for geographically distinct trials by adding the necessary positional dimensions, which also suggests a new approach for data analysis work in big detector arrays.
		
\bibliographystyle{unsrt}
\bibliography{main}	
\end{multicols}
	
\end{document}